\documentclass{ws-jai}

\usepackage[latin9]{inputenc}
\usepackage{graphicx}

\usepackage[breaklinks,colorlinks,urlcolor=blue,citecolor=blue,linkcolor=blue]{hyperref}
\usepackage{url}

\usepackage{subfigure}
\usepackage{color}
\usepackage{threeparttable}

\begin{document}

\catchline{}{}{}{}{} 

\markboth{D. C. Price et. al.}{JAI Introduction}

\title{Introduction to the Special Issue on 
       Digital Signal Processing in Radio Astronomy}  

\author{
D.~C.~Price$^{1\dagger}$, J.~Kocz$^2$, M.~Bailes$^3$ and L.~J.~Greenhill$^4$
}

\address{
\small
$^1$University of California Berkeley, Campbell Hall 339, Berkeley CA 94720\\
$^2$California Institute of Technology, 1200 E. California Blvd, Pasadena, CA, 91125\\
$^3$Swinburne University of Technology, Centre for Astrophysics and Supercomputing, 
\\Mail H39, PO Box 218, VIC 3122, Australia\\
$^4$Harvard-Smithsonian Center for Astrophysics, 60 Garden Street, Cambridge, MA, 02138, USA\\
}

\maketitle
\corres{$^\dagger$Corresponding author. Email: \url{dancpr@berkeley.edu}}


\begin{abstract}
Advances in astronomy are intimately linked to advances in digital signal processing (DSP). This special issue is focused upon advances in DSP within radio astronomy. The trend 
within that community is to use off-the-shelf digital hardware where possible 
and leverage advances in high performance computing. In particular, graphics 
processing units (GPUs) and field programmable gate arrays (FPGAs) are being 
used in place of application-specific circuits (ASICs); high-speed Ethernet and 
Infiniband are being used for interconnect in place of custom backplanes. 
Further, to lower hurdles in digital engineering, communities have designed and 
released general-purpose FPGA-based DSP systems, such as the CASPER ROACH board, 
ASTRON Uniboard and CSIRO Redback board. In this introductory article, we give a brief historical
overview, a summary of recent trends, and provide an outlook on future directions. 

\end{abstract}

\section{Introduction}

Our understanding of the Universe is inextricably linked to technological advances. The instruments we build to probe the nature of our surroundings inform us  about the nature of the cosmos and enable us to test hypotheses and update  theories. Since the 1960s, the computational capability of digital systems has roughly doubled every two years (tracking ``Moore's Law");  consequently, the capabilities of digital instruments continually evolve, allowing better, more sensitive measurements to be conducted.

This special issue on digital signal processing (DSP) in radio astronomy celebrates the capabilities of digital systems for radio astronomy in 2016. The past several years have seen convergence between the fields of high-performance computing (HPC) and radio astronomy. Increasingly, astronomers are able to use off-the-shelf compute hardware and leverage HPC software and frameworks. This has led to faster adoption of new technology and dramatically decreased development timelines. In this introductory article, we give a summary of the current capabilities, trends, and design methodologies within radio astronomy, beginning with a historical overview of how we got here. 

\section{A historical overview}

\subsection{Pioneering steps}

The digital signal processing (DSP) revolution for radio astronomy kicked off in earnest in 1961, with Sander Weinreb applying digital spectral analysis techniques to radio astronomy \citep{Weinreb:1963tn}. Weinreb implemented a 1-bit digital autocorrelation spectrometer, which was then used to make the first detection of the hydroxyl radical (OH) in the interstellar medium \citep{weinreb:1963p9992}. This is particularly remarkable as the Fast Fourier Transform (FFT) algorithm---used in most modern spectrometer systems---was not discovered until 1965\footnote{While Cooley \& Tukey are credited with the development of the general FFT algorithm, Gauss was aware of ``fast'' Fourier Transforms as early as 1805.} \citep{Cooley1965}.

Use of autocorrelators for spectroscopy is a cornerstone of radio astronomy, with bandwidths for modern systems exceeding several GHz.  Another is the use of an array of antenna elements (an interferometer) to increase sensitivity and/or angular resolution. Synthesizing images from radio interferometric data requires computing  the cross correlation of voltage output time series for every pair of antenna receivers. The number of pairs increases quadratically, making large arrays computationally infeasible in the early days of radio astronomy. An alternative approach is to sum the products of calibration factors and voltage outputs, in a process known as "beamforming."  Pointing to each pixel on the sky requires different factors in the summation. Early instruments include Ryle and Vondberg's two-element interferometer (1946), the Cambridge interferometer (1955) and the Mills Cross (1958); further historical details, as well as derivation of technical fundamentals, may be found in \citet{ThompsonMoranSwenson2004}.   

The most scientifically productive such facility to date is the Very Large Array, operated by the National Radio Astronomy Observatory in New Mexico;\footnote{The National Radio Astronomy Observatory is a facility of the National Science Foundation operated under cooperative agreement by Associated Universities, Inc.} the maximum baseline employed there in real-time cross correlation is about 35 km with individual bandwidths of up to 2~GHz sampled at 3 bits.  Extension of array angular resolution for fixed frequency  was made possible by the availability of digital recording systems, resulting in development of Very Long Baseline Interferometery (VLBI), in which signals from radio telescopes separated by hundreds and thousands of kilometers are played back and combined to form synthesized images with greater resolution. \citet{Bare1967} were the first to demonstrate digital VLBI techniques; \citet{Moran1967} demonstrated a VLBI recorder in the same year . These systems recorded a few MHz of bandwidth with 1-bit sampling. Modern VLBI recorders  \citep{Whitney2010,Vertatschitsch2015} are able to record several GHz of bandwidth, and generally store data reduced to 2-bits.  Theory and technique are developed in depth by \citet{ThompsonMoranSwenson2004}.   
%
%
%
%

\subsection{The Rise of FX correlators}

\begin{figure}[t]
 \centering
 \includegraphics[width=0.9\textwidth]{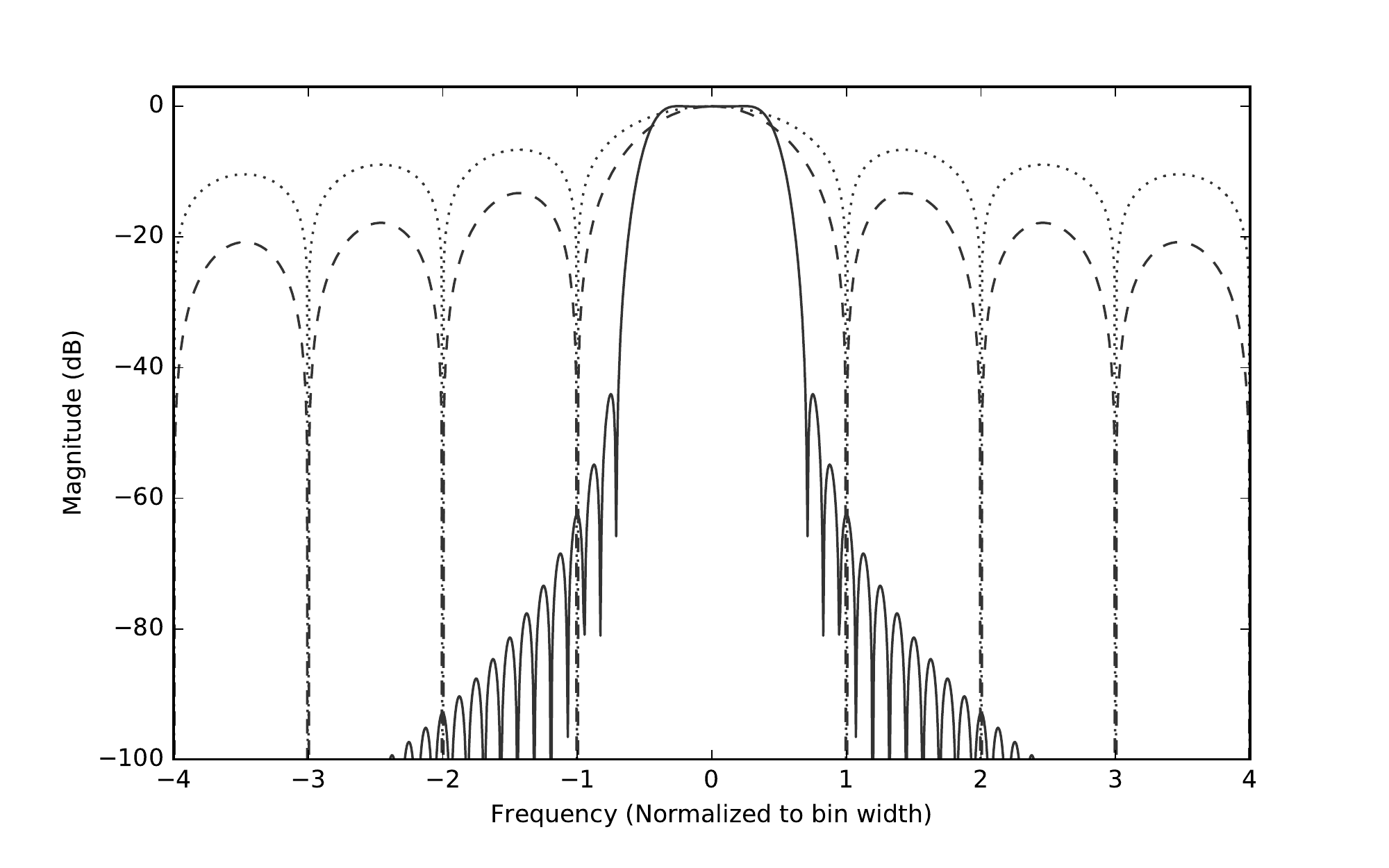}
 \caption{Comparison of the filter response for an XF (dotted), FX (dashed) and an 8-tap, Hamming-windowed PFB (solid). The PFB structure greatly increases inter-channel isolation. While windowing functions may be applied to the XF and FX data, this increases the width of the filter. \label{fig:pfb}}
\end{figure}

\begin{figure}[t]
 \centering
 \includegraphics[width=0.9\textwidth]{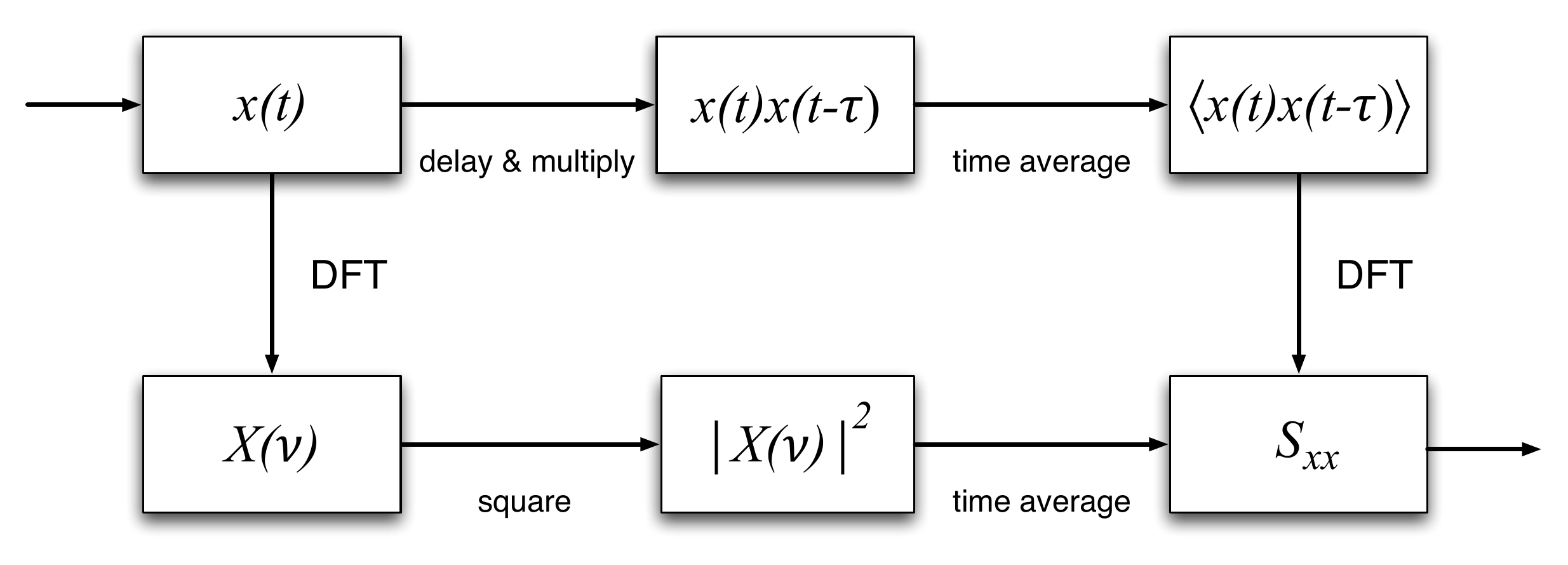}
 \caption{Schematic representation of the Wiener-Khinchin theorem, which states that a power spectrum, $S_{xx}$, may be constructed from a time series of measurements, electric field $x(t)$, via transformation in the time or frequency domains (the upper or lower paths, respectively).  \label{fig:wiener}  }
\end{figure}

The first correlators used an `XF' implementation, in which the delay-and-multiply correlation function of a signal pair is calculated (X), the data are averaged, and then a discrete Fourier transform is applied (F). In contrast, most modern-day systems use an `FX' implementation, in which a discrete Fourier transform is applied (F) before cross multiplication of signal pairs (X).   The order of operations is important: the two approaches are related by the well-known Wiener-Khinchin theorem (Figure~\ref{fig:wiener}), but are not equivalent (e.g.,\citet{Price2016}). 
Use of the highly-efficient FFT algorithm makes the FX approach significantly less computationally intensive than the XF approach. For even a modest number of channels, e.g., $N_{\rm{chan}}$=1024, an FX correlator requires 0.1\% the computation required for a comparable XF implementation \citep{ThompsonMoranSwenson2004}. 

Given the significant computational saving afforded by FX architectures, one may wonder why XF architectures were commonly used. There are several reasons. Firstly, initial correlator designs preceded publication of  the FFT algorithm. Secondly, one-bit multiplication can be performed using a simple AND logic gate, and low-bitwidth multiplier circuits are similarly straightforward to implement in hardware; further, before the late 1970s, DSP chips lacked a multiply instruction. There being significant quantization errors when low bitwidths were used within the multiple computational stages of an FFT introduced additional complications.  The first published use of an FX system within radio astronomy can be found in \citet{Chikada:1987p10044}, who implemented a 1024-channel system for the Nobeyama Radio Observatory c.1983. Almost all recently developed correlator and spectrometer systems use the FX architecture, and in view of modern technology, use of the XF architecture will probably be limited to special use cases, e.g.,  arrays with few baselines. 

\subsection{The Rise of Polyphase Filterbanks}

A polyphase filterbank (PFB) is a computationally efficient structure used to create a band of digital filters. This is constructed from a polyphase finite impulse response (FIR)  filter frontend that precedes an FFT; it offers far better isolation between channels than both XF and FX implementations (Figure~\ref{fig:pfb}). PFB-based correlators can still be considered 'FX' as the FFT can itself be considered a critically-sampled filterbank. The PFB architecture was first introduced by \citet{Schafer1973}, and further popularized by \citet{Bellanger:1976p7898}. 

Despite their advantages, the first published use of a PFB in radio astronomy was in 1991 \citep{zimmerman1991}, where a PFB-based spectrometer was commissioned for SETI (Search for Extraterrestrial Intelligence) observations. It was not until 2000 that PFBs were integrated into cross-correlation systems \citep{Bunton2000}. Recently developed radio interferometers, such as ASKAP \cite{DeBoer2009}, MeerKAT \citep{Jonas2009}, LOFAR \citep{deVos2009}, and LEDA \citep{Kocz2015}, use PFB-based correlators. The increasing time and frequency occupancy of the radio spectrum due to broadcasting, sensing, and ever denser advanced infrastructure, and the corresponding challenges of mitigating radio frequency interference (RFI) have been major drivers in the adoption of  PFB-based systems. 

\section{Recent trends}

\subsection{Heterogeneous Systems}

Increasingly, radio astronomers leverage off-the-shelf commodity hardware connected together by industry standard "interconnects," commonly high-speed Ethernet and Infiniband. A now-common design pattern is to divide DSP tasks up between a "frontend" that does low-level DSP (e.g. digitization and channelization) and a "backend" that performs higher-level DSP tasks that are more complex, are most likely to evolve or involve reconfiguration, or are more efficiently executed on a different architecture than that used for the frontend.     Systems that mix DSP architectures are referred to as \emph{heterogeneous} systems, as in High Performance Computing (HPC).

Several heterogeneous systems are detailed within this special issue: 
\begin{itemize}
	\item The HI-Pulsar system (HIPSR), a spectrometer system for the Parkes multibeam receiver \citep{Price2016b}.
	\item The Canadian Hydrogen Intensity Mapping Experiment (CHIME); the custom interconnect between the frontend and backend is detailed in this special issue \citep{Bandura2016b}.
	\item The Deep Space Network's transient observatory digital systems \citep{Kuiper2016}, 
\end{itemize}
These systems all use Field-Programmable Gate Array (FPGA) processing boards for frontend computation, coupled with backends consisting of high-performance compute servers equipped with Graphics Processing Units (GPUs).  Each platform has particular strengths.  For FPGAs, these include synchronous operation, reconfigurable high-speed bus communication, and resource-efficient fixed precision arithmetic.   For GPUs, these include massively parallel execution, transparent and dynamic scheduling of processing threads,  exceptional compute density, high-efficiency compilers, and sometimes less onerous programming and debug environments. 

\subsubsection{Correlation}
The first digital correlators were custom systems, built from scratch for their very specific purpose.  Upgrades to bandwidth or numbers of antennas beyond initial specifications could require wholesale system re-engineering or replacement (e.g., the recently completed upgrade to the VLA \citep{perley:2009fp}).  Heterogeneous computing clusters dedicated to correlation combine FPGAs, for channelization and/or digitization, and GPUs for cross-multiplication and related tasks, such as thresholding, high-order moment calculation for RFI detection, time gating, and beamforming that occur before time averaging.   FPGA-GPU correlators have become common in low-frequency radio astronomy, including the CHIME \citep{Vanderlinde2014}, LEDA \citep{Kocz2015},  Murchison Widefield Array \citep[MWA, ][]{Ord2015}, Hydrogen Epoch of Reionization Array \citep[HERA, ][]{DeBoer2016} systems, among others.  While it has been speculated that full-cost, full life-cycle accounting for development, construction, operations, and upgrades favors systems that leverage GPUs, as opposed to FPGAs exclusively, there has not been a comprehensive analysis, and thus far, FPGA-GPU systems have catered to  experiments and smaller facilities.  
 
%
%
\subsubsection{Pulsars}
Current pulsar observing systems can be broadly broken into two categories: survey/search and timing. Pulsar survey data are digitized using FPGA-based components and data are processed via a polyphase filterbank, accumulated to a select time resolution, and stored for off-line processing.  In timing studies, digitized data are passed to either a CPU or GPU-based system that coherently de-disperses in time and frequency, and folds the data to obtain a pulse profile.  Recently implemented searching and timing pipelines \cite{Kuiper2016,Kocz2016} are described in this issue. 
%
%
%
%
While the data from pulsar surveys are stored for offline processing, search pipelines executing on GPUs have enabled real-time detection of some signals. This has been particularly useful for the detection of rotating radio transients \citep[RRATs][]{mcLaughlin2005} and fast radio bursts \citep[FRBs][]{lorimer2007}. Rather than looking for periodic signals, GPU based software such as \texttt{HEIMDALL}\footnote{\url{https://sourceforge.net/projects/heimdall-astro/}} and \texttt{astro-accelerate}\footnote{\url{https://github.com/wesarmour/astro-accelerate}} is used to search the filterbank for strong single pulses. Rapid follow up on these signals can then be conducted, or raw data stored for more detailed analysis. A perennial challenge for such systems can be false positives generated as a result of RFI.   Real-time filtering algorithms \citep[e.g. ][]{Dumez2016} to differentiate true sources from RFI can be necessary, to ensure low time and frequency occupancy of interference, and to achieve high detection efficiency depending on circumstances.  
%
%
%
%
%
\subsection{MKID readout systems}

Microwave Kinetic Inductance Detectors (MKIDs) are a new type of detector and alternative to bolometer arrays. First developed by \citet{Day2003}, MKIDs are superconducting resonator circuits whose resonant frequency changes when an incident photon is absorbed. These circuits are readily multiplexed, allowing thousands of detectors to be read out over a single transmission line. In comparison to competing bolometer technologies, MKID systems essentially transfer the complexity of multiplexing from cryogenic electronics into room-temperature electronics and signal processing systems \citep{mchugh2012}. Among other advantages, MKIDs promise to make construction of arrays with many tens of thousands of detectors tractable. 

MKID readout systems require digital to analog converters (DAC) that are matched in bandwidth and dynamic range to a high-bitwidth analog to digital converter (ADC). A comb of probe frequencies are generated by the DAC at the resonant frequencies of the MKID array; the ADC digitizes the output of the MKID.  Changes in the resonators' amplitude and phase are monitored via the use of a PFB. While MKIDs typically operate at near infrared to ultraviolet wavelengths, MKID readout systems share many similarities in hardware and firmware to radio astronomy spectrometer systems; consider use of the open-source hardware \citep{Hickish2016} from the Collaboration of Astronomy and Signal Processing (CASPER\footnote{\url{https://casper.berkeley.edu}}) for both spectrometer and MKID backends. The CASPER-based BLAST-TNG system \citep{Gordon2016}, detailed in this issue, provides a good example of a MKID readout system.
%
%
%
%

\section{Current technologies}

\subsection{Analog to Digital Converter}

ADCs are characterized, grossly, by sampling speed and bit depth. For radio astronomy applications, where signals are noise-like, ADCs of only a few bits are sufficient. However, the presence of RFI motivates the use of 8-bit and higher ADCs with larger dynamic range. As an example, the DSPZ digital baseband receiver system \citep{Zakharenko2016} uses a 16-bit ADC to achieve a $\sim96$ dB dynamic range (in power) and enable operation within the heavily contaminated 8-32~MHz band.

In an ideal system design, the entire usable bandwidth of the receivers is digitized for downstream processing and science. For example, the SWARM correlator \citep{Primiani2016} processes 32~GHz of bandwidth provided by the wideband receivers of the Smithsonian Submillimeter Array (SMA). SWARM uses a set of 8-bit, 5 Gsample/s digitizer cards (CASPER ADC1x5000-8), each of which processes a 2~GHz wide band mixed down to baseband. Wider bandwidth ADCs, such as the Analog Devices 3-bit, 26 Gsample/s part\footnote{HMCAD5831, \url{http://www.analog.com/media/en/technical-documentation/data-sheets/hmcad5831.pdf}}, and the Vadatech 8-bit, 56 Gsample/s board\footnote{AMC590, \url{http://www.vadatech.com/product.php?product=404}}, are commercially available but not yet in use within radio astronomy.  Where science drivers do not demand such high bandwidths, sampler resources can be traded off to provide a high density of signal inputs, e.g., the Hitite HMCAD1511\footnote{HMCAD1511, \url{http://www.analog.com/media/en/technical-documentation/data-sheets/hmcad1511.pdf}} used in the LEDA project where 512 receiver signals paths are condensed for digitization by just 16 FPGA nodes \citep{Kocz2015}.

\subsection{Field-Programmable Gate Array}

FPGAs are reconfigurable integrated circuits that consist of an array of programmable logic blocks and a reconfigurable interconnect. The interconnect allows for logic blocks to be "wired" together to make a digital circuit for a specific application. FPGAs excel at low-level DSP tasks with high data throughput requirements and are easily interfaced with ADCs, DACs, and interconnects.  In order to be useful for radio astronomy, FPGAs must be embedded on a board with peripheral interfaces. CASPER  provided the radio astronomy community with open-source FPGA-based hardware for over a decade. Several systems detailed in this special issue are based upon the Xilinx Virtex-6 based ROACH-2 processing platform; see  \citet{Hickish2016} for a comprehensive overview. Also in this issue, the Xilinx Kintex-7 based ICE FPGA platform --- developed by the McGill Cosmology Instrumentation Laboratory  --- is introduced \citep{Bandura2016a}. Several other FPGA-based processing boards designed specifically for radio astronomy exist: ASTRON has developed the Uniboard platform \citep[Intel Stratix IV FPGA,][]{Szomoru2010}, and CSIRO have developed the Redback platform \citep[Xilinx Kintex-7,][]{Hampson2014}. NetFPGA\footnote{\url{http://netfpga.org/}} provide general-purpose FPGA boards, such as the Virtex-7 based SUME. Detailed in this issue is firmware for real-time RFI mitigation \citep{Dumez2016} that runs upon the Uniboard platform. 

The current state-of-the-art FPGAs are the Xilinx Ultrascale+ series and Intel Stratix 10 (formerly Altera). While traditionally FPGAs implement fixed-point arithmetic, both Intel and Xilinx FPGAs now support IEEE 754-compliant floating point (FP) operations, with Stratix 10 chips offering dedicated single-precision DSP blocks. While HDL remains the standard language used for FPGA programming, the OpenCL standard\footnote{\url{https://www.khronos.org/opencl/}} is now supported by both vendors.

\subsection{Graphics Processing Units}

In recent years, GPUs have become prevalent in radio astronomy, not just in DSP.   GPUs are massively parallel compute engines.  Discrete units typically have O(1000) "cores" that are served by a high speed hierarchical memory stack included in which are registers and shared memory on-die and O(10 GB) DDR5, or faster, RAM off-die.  This enables compute intensive, parallelizable algorithms to be offloaded from central processing units (CPU), which have parallelized vector pipelines but are chiefly optimized for serial operations in general computing and provide O(10) cores.  GPUs are programmed using either NVIDIA CUDA\footnote{\url{https://developer.nvidia.com/cuda-zone}} or OpenCL, both of which augment C/C++ with GPU-specific functionality.  

GPU efficiency may be measured by different code-dependent metrics.   Resource utilization refers to the fraction of the theoretical compute capacity, excluding  I/O usage, accessed by an algorithmic implementation.  \citet{Clark:2013fr} achieved 79\% single precision resource utilization for cross-correlation using the xGPU kernel.\footnote{https://github.com/GPU-correlators/xGPU}.  Power efficiency (e.g., 32-bit FP operations per Watt) is a gross figure that includes the costs of data transport as well as computation by a subset of available resources).  Power consumption has been regarded as a drawback of GPUs for some applications, but even for those that are not compute bound, power may not be a critical system engineering consideration (cf. execution time and scalability) 

That said, the recent advent of fixed-point and low-bitwidth instructions in off-the-shelf GPU hardware (8 and 16-bit), driven by visualization and Deep Learning applications in other disciplines, has the potential to boost power efficiency considerably for radio astronomical applications involving noise-like signals.  This includes pulsar and transient pipelines \citep{Barsdell2012, adamek2016b}, polyphase filterbanks \citep{chennamangalam2014, adamek2016},  and for correlator X-engines in heterogeneous systems \citep{harris2008, Clark:2013fr}.   The NVIDIA Pascal (2016) architecture was the first to offer both 8 and 16-bit instructions that are accelerated with respect to 32-bit floating point.  

Three papers in this special issue utilize GPUs: the HIPSR system at Parkes \citep{Price2016b}, the AARTFAAC all-sky monitor \citep{Prasad2016}, and the pulsar timing system at the Deep Space Network \citep{Kocz2016}. 

\begin{table}
\begin{center}
\begin{threeparttable}
	\caption{Specifications of selected correlators with large  figures of merit, $N_{\rm{ant}}B$ and $N_{\rm{ant}}^2B$. Note that this table does not account for other characteristics, such as  beamforming capability, data output rate, etc. \label{tab:corr}}
	\begin{tabular}{l r c c c r}
	\hline
	Telescope	& Reference & $N_{\rm{ant}}$ &  $B$ & $N_{\rm{ant}}B$ & $N_{\rm{ant}}^2B$ \\
		& &                & (GHz)  & (GHz)             & (GHz)         \\
    \hline
	CHIME-1024 		 & \citet{Vanderlinde2014}  & \textbf{1024} & 0.4    & 409.6  & \textbf{419430} \\ 
	ALMA	 		 & \citet{Baudry2012}   & 64	   & \textbf{16.0}	    & \textbf{1024.0}  & 65536 \\
	HERA-352         & \citet{DeBoer2016}       & 352  & 0.2    & 70.4   & 24781 \\
	ASKAP		 	 & \citet{Tuthill2014}       & 36   & 0.3  	& 388.8$^\dagger$	& 13997$^\ddagger$ \\ 
	eVLA	 		 & \citet{perley:2009fp}    & 27	   & 8.0	    & 216.0  & 5832 \\
	LEDA	 		 & \citet{Kocz2015}         & 256  & 0.058  & 14.85	 & 3801 \\
	MeerKAT		     & \citet{Jonas2009}        & 64	   & 0.856	& 54.78	 & 3506 \\
	AARTFAAC-12	     & \citet{Prasad2016}       & 576  & 6.25   & 3.6    & 2074 \\
	PAPER-128		 & \citet{Cheng2016}          & 128	  & 0.100   & 12.8	    & 1638 \\
	SMA				 & \citet{Primiani2016}     & 8    & \textbf{16.0}   & 128.0  & 1024 \\
	MWA				 & \citet{Ord2015} & 128  &	0.030   & 3.84	& 492 \\
	uGMRT            & \citet{reddy2016}           & 32   & 0.4   & 12.8 & 410 \\
    EOVSA            & \citet{nita2016}           & 16   & 0.6   & 96 & 154 \\
	LOFAR	         & \citet{deVos2009}          & 48	  & 0.032  & 1.54   & 74 \\
	\hline
	\end{tabular}
	
 \begin{tablenotes}
 \item [$\dagger$] Computed as $N_{\rm{beam}}N_{\rm{ant}}B$, with the number of beams $N_{\rm{beam}}$=36.
 \item [$\ddagger$] Computed as $N_{\rm{beam}}N^2_{\rm{ant}}B$ with $N_{\rm{beam}}$=36.
 \end{tablenotes}

\end{threeparttable}
\end{center}
\end{table}

\section{Future outlooks}

\subsection{Drivers for correlators}

While correlators are only one of many types of DSP backend, their design, size and implementation platform are indicative of the technology available during their design period. Two main factors driving the requirements of cross-correlators are the number of antennas, $N_{\rm{ant}}$, and the processed bandwidth, $B$. At low-frequency (under 1~GHz), the ratio of bandwidth to $N_{\rm{ant}}$ is low, whereas at high-frequencies (above 100~GHz), the ratio of bandwidth to antennas high. For example, the correlator for the Long Wavelength Array (LWA) at Owens Valley has $N_{\rm{ant}}$=256 and $B$=58~MHz \citep{Kocz2015}, whereas the Atacama Large Millimeter Array \citep[ALMA,][]{Wootten:2009hw} has $N_{\rm{ant}}$=66 and $B$=8~GHz. 

For cross-correlators, two figures of merit for digital systems are their bandwidth and compute requirements. The aggregate data rate or throughput is simply $N_{\rm{ant}}B$, and the number of computations required scales as $N^2_{\rm{ant}}B$. Table~\ref{tab:corr} lists the `largest' correlators in the world, defined by  $N^2_{\rm{ant}}B$.  It should be noted that these two figures of merit do not fully reflect the complexity or capability of every system.
  

\subsubsection{Large-$N_{\rm{ant}}$, low-frequency arrays}

In general, increasing the number of antennas in an interferometric array increases the collecting area and imaging capabilities, including optimization of the interferometer point spread function.  Adding longer baselines increases imaging angular resolution.  The number of computations required by an interferometer scales proportionally to $N_{\rm{ant}}^2B$, making the addition of antennas computationally expensive. At low frequencies, arrays of many hundreds of elements are now commonplace --- for example, the Long Wavelength Array \citep[LWA][]{Ellingson:2009p531}, the Murchison Widefield Array \citep[MWA,][]{Lonsdale:2009p2402} and the Low Frequency Array \citep[LOFAR,][]{deVos2009}. 

So-called $\lambda21$-cm cosmology \citep{Pritchard2012} is currently the main science driver for  ``large-$N_{\rm ant}$''  correlators and the DSP challenges they pose.  Of particular importance for these arrays is tailoring of field of view (trading off element size and number) and coverage of the 2D plane of baseline spacings (e.g., a fully-filled aperture or one that redundantly samples particular angular scales). The Canadian Hydrogen Intensity Mapping Experiment, CHIME \citep{Vanderlinde2014}, seeks to push to $N_{\rm{ant}}$=1024, with a bandwidth of $B$=400~MHz. To accomplish this, CHIME have developed the low-cost FPGA platform, ICE \citep{Bandura2016a}, implemented a custom interconnect using the ICE platform \citep{Bandura2016b}, and will use GPUs for cross correlation \citep[i.e., the X-engine,][]{Denman2015}. The Hydrogen Epoch of Reionization Array \citep[HERA,][]{DeBoer2016}, is employing a similar large-$N_{\rm{ant}}$ approach, with $B$=200~MHz and $N_{\rm{ant}}$=352.

The Square Kilometre Array \citep[SKA,][]{Dewdney:2009p1690,Dewdney2015} seeks to dramatically increase $N_{\rm{ant}}$ to $\sim$130,000 in a Phase 1 low-frequency aperture array system when built (SKA-low). Also driven primarily by $\lambda 21$-cm cosmology, SKA-low will consist of $\sim$512 stations, each with 256 antennas that are beamformed. While the total number of antennas is larger than CHIME, the total number of beamformed inputs to the correlator will be fewer.

\subsubsection{Next-generation dish arrays}

High frequency and angular resolution line surveys at frequencies of 1-10~GHz over a large redshift range (to enable statistical cosmological studies), and deep pulsar surveys, are two of the myriad science drivers for the anticipated SKA mid-frequency array that will drive DSP specifications.  This will be built out from the current 64-dish MeerKAT array, to a total  $N_{\rm{ant}}$=197 and target bandwidth of $B$=770~MHz \citep{Dewdney2015}.  While this is indeed a large correlator, its processed data rate ($N_{\rm{ant}}B$) will be about a third that of the ALMA, and the compute requirements ($N^2_{\rm{ant}}B$) are roughly the same. While other factors, e.g., number of PFB channels, data output rate, numbers of processing backends, make a detailed comparison less straightforward, the SKA-mid correlator is eminently realizable. 

The `Next-Generation Very Large Array' (ngVLA\footnote{\url{https://science.nrao.edu/futures/ngvla}}) is a second large dish-array concept now in formulation.  The array will likely target a frequency range between the SKA-mid and ALMA millimeter bands with some overlap (a few cm to a few mm), with a collecting area of $\sim10$ times that of the extant array \citep[VLA,][]{perley:2009fp}, provided by $\sim300$ antennas, with O(10)~GHz instantaneous bandwidth.  High continuum sensitivity, high angular and frequency resolution, full beam mapping capability, and fast spectral line survey speeds are likely to be critical system design parameters driving DSP requirements.


\subsection{Single-dish telescopes}

The digital backend requirements for single-dish telescopes are driven by the characteristics of the receiver: in particular by the usable bandwidth, the number of beams delivered, and the required time resolution of output data products. While the collecting area of a dish is fixed, observational capabilities can nonetheless be enhanced via design and installation of improved receivers. 

Phased array feeds --- feeds in which an array of antenna elements are phased together to form directional beams --- are a relatively new technology that demands higher bandwidth and computational capacity from the DSP backend. Studies and development efforts for phased array feeds  \citep[e.g. ][]{CortesMedellin:2015bp, Chippendale:2016br} are underway at most major single dish facilities.

The development of ultra-wideband feeds with greater than 6:1 fractional bandwidths \citep[e.g. ][]{Jacobs:2013gt, Yin:2013je} is  driving the requirement for DSP systems capable of ingesting wider bandwidths. Such feeds promise to greatly improve the spectral survey speed and continuum sensitivity of the telescope, provided capable DSP systems are developed in tandem.

Two significant single-dish telescopes have recently been constructed: the partially steerable Five-hundred meter Aperture Spherical Telescope \citep[FAST,][]{li2012, fast_sciencemag} and the fully steerable 64-m Sardinia Radio Telescope \citep[SRT,][]{bolli2015}. Not yet 20 years old, largest fully-steerable telescope in the world, the 110-m aperture Robert C. Byrd Green Bank Telescope (GBT) has already served a number of DSP development efforts advancing spectroscopy and pulsar investigations. What appears to be premature divestment by the NSF \citep{wired_gbt}, judging from the facility science output however, suggests an uncertain future. Nonetheless the uniqueness of this ``science-ready,'' well characterized telescope (e.g., active surface, $\sim 100$~GHz frequency range, off axis structure and exceptional sensitivity, and siting in the National Radio Quiet Zone) make it an exceptional platform for next-generation DSP development and a lynch pin for frontier pulsar timing and gravity wave science \citep{lockman2016}.


\subsection{Realtime data reduction}

With ever-increasing data rates comes pressure to conduct data reduction in real time: writing data to high-capacity, long-term media remains a significant bottleneck. The stark challenge of realtime data reduction is well illustrated by the SKA, which will produce approximately 10 exabytes per day into status buffer memory \citep{quinn2015}. At such tremendous data volume, it is imperative that data reduction is performed in realtime. 

This special issue highlights several examples of tasks traditionally considered post-processing becoming pre- or real-time processing: pre-correlation RFI flagging \citep{Dumez2016}, RFI mitigation at the upgraded Giant Meter Wave Radio Telescope -- uGMRT \citep{buch2016}, the AARTFAAC all-sky monitor \citep{Prasad2016}, and real-time fast radio burst detection in the HIPSR system \citep{Price2016b}.

\subsection{Analog Digital Converters}

In terms of ADC interfaces, the JESD204B standard\footnote{\url{http://www.jedec.org/standards-documents/results/jesd204}} is steadily gaining momentum over LVDS and CMOS. JESD204B allows the use of multi-gigabit transceivers, meaning a reduction in interface pin count over LVDS. Many next-generation ADCs utilize JESD204B, and it is likely that this standard will become more and more common as ADC bandwidths continue to grow.

\section{Conclusions}

The world of DSP evolves quickly. New technologies and increasingly performant hardware are continually developed, ``obsoleting'' existing systems. For radio astronomy, we may expect these innovations to lead to construction of larger arrays of antennas, wider instantaneous bandwidths, and more computationally expensive data reduction and search algorithms running in real time. 

Concomitant adaptation of algorithms to new platforms quickly and effectively will be critically important to realizing this future.  Arguably, even more important will be timely community publication and dissemination of the details of DSP implementations and advances made through application of new techniques and novel technologies. Indeed, this is a key motivation for the special issue. 
\bibliographystyle{ws-jai}
\bibliography{journals,jai-intro}

\end{document}